\documentstyle[12pt]{article}
\begin{document}
\baselineskip24pt
\begin{center}
{\Large{\bf DUALITY AND MASSIVE GAUGE INVARIANT THEORIES}}
\vskip0.5cm
E.Harikumar and M. Sivakumar$^{**}$
\vskip0.5cm
{\it School of Physics, University of Hyderabad,\\
Hyderabad - 500 046, {\bf India}}
\end{center}

\begin{center}
{\bf ABSTRACT}
\end{center}

Two different massive gauge invariant spin-one theories in $3+1$
dimensions, one Stuckelberg formulation and the other
`$B^{\wedge}F$' theory, with Kalb-Ramond field are shown to be
related by duality. This is demonstrated by gauging the global
symmetry in the model and constraining the corresponding dual
field strength to be zero by a Lagrange multiplier, which
becomes a field in the dual theory. Implication of this
equivalence to the  $5$ dimensional theories from which these
theories can be obtained is discussed. The self-dual
Deser-Jackiw model in $2+1$ dimensions, is also shown to result
by applying this procedure to Maxwell-Chern-Simon theory.

\vfill
${**}$ email: mssp@uohyd.ernet.in
\newpage

Massive gauge invariant spin-one theories has been studied for a
long time with two principle procedure for it: Schwinger
mechanism in $1+1$ dimension [1] of $2$d quantum electrodynamics
of massless fermion, yielding massive gauge field through axial
anomaly and Higgs mechanism. In $3+1$ dimensions massive
spin-one theories are generally considered following one of the
two procedures: one by Stuckelberg formulation [2] and the other
by using Kalb-Ramond field (rank two antisymmetric tensor gauge
field) [3] in a Chern-Simons like formulation known as `$B^{\wedge}F$'
theory [4]. The latter is well studied in different contexts,
including as realizations of certain condensed matter system [5]
by Sodano {\it et al.,} as an alternate to Higgs mechanism [6]
and as realization of Bosonised Schwinger model in $3+1$
dimensions by Aurilia and Takahashi [7].  On the other hand, the
Stuckelberg formulation of spin-one, (and also for higher spin
fields) [8] have been studied in various contexts, like for
consistency problems in higher spin fields [9] and in string
field theory as description of massive modes [10]. Though they
appear as different construction for maintaining gauge
invariance in the presence of mass terms, in this paper we show
that the two theories are related by duality transformation.

In recent times duality is being studied in detail, due to
developments in string theory, where different inequivalent
vacua are shown to be related by duality [10]. In the context of
sigma models, a procedure for constructing dual theory was given
by Busher [12] and generalized by Rocek and Verlinde[13].
Basically, the procedure consists in gauging the global symmetry
with gauge fields, whose field strength is constrained to be
zero by means of a Lagrange multiplier. Integrating the
multiplier field and then the gauge field, original action was
recovered. Instead, if one integrates the original and gauge
fields, keeping the multiplier field, dual theory was obtained.
This procedure of obtaining dual theory was applied to $1+1$
dimensional Dirac theory and Bosanisation rules were obtained
from this duality procedure by gauging the global phase symmetry
by Burgess and Quvedo [14]. This procedure has also been applied
to gauge theories and shown to lead to S-duality, which relates
strong and weak coupling [15]. This procedure has also been
shown to be related to canonical transformations [16].

In this paper, we apply this procedure of dualization to
toplogically massive gauge theories. We first apply this method
to $B^{\wedge}F$ theory in $3+1$ dimensions where the global
symmetry is shift of the field and show that Stuckelberg type
massive theory is obtained. First we note that, in the former
theory, current due to local gauge symmetry is conserved as an
algebraic identity, like that for topological currents and in
the latter case it is conserved due to the equation of motion
for Stuckelberg field. Since this interchange between
topological and Noether current, generally takes place under
duality transformation, it is plausible that the two theories
are related by duality transformation. This is demonstrated in
this paper by the procedure outlined above. Next topologically
massive $2+1$ dimensional Maxwell-Chern-Simons theory is
considered and this procedure is shown to lead to Deser-Jackiw
model [17] of self dual massive theory.  This paper is organized
as follows:

Section 1 deals with equivalence between `$B^{\wedge}F$' theory
and Stuckelberg theory. Section 2 applies it to
Maxwell-Chern-Simons theory. Finally we end with discussion.
Both these models, {\it i.e.}, `$B^{\wedge}F$' model and
Stuckelberg model has earlier [18,19]  been shown to be obtained
from different $5d$-theories. This result, in the context of
$5d$-theories is also discussed, in the final section.

We use the metric $g_{\mu \nu} = diag(1, -1, -1, -1)$ and
$\epsilon_{0 1 2 3}=1$
\newpage
Section I:

Consider the Lagrangian for topologically massive spin-one theory 
involving Kalb-Romand field $B_{\mu \nu}$ and a vector field $A_\mu$ 
known as $B^{\wedge}F$ theory.
\begin{equation}
{\sl L} = - \frac{1}{4} F_{\mu \nu}^2 + \frac{1}{2 (3!)} 
H_{\mu  \nu  \lambda}^2
- \frac{m}{3!} H_{\mu  \nu  \lambda} \epsilon^{\mu  \nu  \lambda  \rho} 
A_\rho \,\,\,.
\end{equation}
This Lagrangian has local invariance under 
\begin{eqnarray}
A_\mu & \rightarrow & A_\mu + \partial_\mu \Lambda \\
B_{\mu  \nu} &\rightarrow& B_{\mu  \nu} + (\partial_\mu \Lambda_\nu 
- \partial_\nu \Lambda_\mu)\,\,\,.
\end{eqnarray}
where $H_{\mu \nu \lambda} =\partial_ \mu B_{\nu \lambda} + \partial_\nu
B_{\lambda \mu} + \partial_\lambda B_{\nu \mu}$

The field equations following from this  are
\begin{eqnarray}
\partial_\mu F^{\mu \nu} &=& J^\nu \\
\partial_\mu H^{\mu \nu \lambda} &=& J^{\nu \lambda}
\end{eqnarray}
where $J^\mu = \frac{m}{3!} \epsilon^{\mu \nu \lambda \rho}
H_{\nu \lambda \rho}$
and $ j^{\mu \nu} = \frac{m}{3!} \epsilon^{\mu \nu \lambda \rho} 
F_{\lambda \rho}$ are the currents associated with local gauge symmetry 
(2), (3). 

Note that both the currents are conserved as an algebraic identity, like
that of topological current. The fact that this describes massive spin-one
theory can be shown easily by solving the coupled differential 
equations [7].

Next we consider the Stuckelberg formulation of massive vector theory 
whose Lagrangian is
\begin{equation}
{\sl L}= -\frac{1}{4} F_{\mu \nu}^2 - \frac{1}{2} (\partial_\mu \Phi - 
m A_\mu)^2 
\end{equation}
This has invariance under 
\begin{eqnarray}
A_\mu &\rightarrow & A_\mu + \partial_\mu \Lambda \\
\Phi &\rightarrow & \Phi + m \Lambda
\end{eqnarray}
The equation of motion following from this Lagrangian (6) for $A_\mu$
and $\Phi$ are
\begin{eqnarray}
\partial_\mu F^{\mu \nu} +  K^\nu  & = & 0 \\
\partial^\mu (\partial_\mu \Phi - m A_\mu) & = & 0\,\,\,.
\end{eqnarray}
where $K^\nu \equiv m (\partial^\nu \Phi - m A^\nu)$.
 Now note that the current associated with $A_\mu$ field is
conserved due to the equation of motion of $\Phi$ field, like
that of Noether current. The fact that this describes massive
spin-one theory can be seen by using the gauge invariance (8)
and fixing the field $\Phi$ to zero.

Thus we have two (apparently) different formulation of spin-one
theory. But the nature of currents in the two theories and
physical equivalence of the system they describe, {\it viz.,}
massive spin-one particle forces one to enquire if both these
formulations are related by duality transformation. We next
show, indeed that is the case.

The dual theory is obtained by the procedure of gauging the
global symmetry in the model by a gauge field and constraining
its dual field strength to be zero by means of a Lagrange
multiplier and by integrating the original and the gauge field
and expressing the theory in terms of the multiplier field, the
dual theory is obtained. The global symmetry, in question in
this model (1) is $\delta B_{\mu \nu} = \epsilon_{\mu \nu}$ ,
$\delta A_\mu = 0$ (Note by dropping a surface term, the global
symmetry is on the vector field.  This is discussed later). This
symmetry is gauged by introducing a three form gauge potential,
$G_{\mu \nu \lambda}$ in the Lagrangian (1), whose dual field
strength is $\epsilon^{\mu \nu \lambda \rho} \partial_\mu G_{\nu
\lambda \rho}$, which is gauge invariant under $\delta G_{\mu
\nu
\lambda} = \partial_\mu \eta_{\nu \lambda}$. By adding a scalar field 
as a Lagrange multiplier, its dual field strength is constrained to
be flat. The Lagrangian is
\begin{equation}
{\sl L} = -\frac{1}{4} F_{\mu \nu}^2  + \frac{1}{2 (3!)}
(H_{\mu \nu \lambda} - 
 G_{\mu \nu \lambda})^2 - \frac{m}{3!} (H_{\mu \nu \lambda} - 
 G_{\mu \nu \lambda}) \epsilon^{\mu \nu \lambda \rho} A_\rho 
+ \frac{1}{3!} \Phi \epsilon^{\mu \nu \lambda \rho}
\partial_\mu G_{\nu \lambda \rho} \,\,\,.
\end{equation}
Note that the original gauge invariance of the vector theory, 
$\delta A_\mu = \partial_\mu \Lambda $, is 
recovered only when, under $A_\mu$ gauge transformation, scalar field also 
transforms 
\begin{equation}
\Phi \rightarrow \Phi + m \Lambda\,\,\,.
\end{equation}
This transformation is the same as that of Stuckelberg formulation of the 
theory. Indeed, by integrating over $B_{\mu \nu}$ and $A_{\mu}$ fields, 
which appear as  Gaussian, Stuckelberg theory (6) results.

Instead of considering the global symmetry in $2$-form $B$ field, one could
start, after omitting a surface term in the Lagrangian (1),
which has a global symmetry in $A_\mu$ field of the form 
$\delta A_\mu = \epsilon_\mu$ and $\delta B_{\mu \nu} = 0$. Gauging this 
symmetry, one gets
\begin{equation}
{\sl L} = -\frac{1}{4}(F_{\mu \nu}-G_{\mu \nu})^2  +
\frac{1}{2 (3!)} H_{\mu \nu \lambda}^2  +
\frac{1}{2} m B_{\mu \nu } \epsilon^{\mu \nu \lambda \rho}
(F_{\lambda \rho}-G_{\lambda \rho})
+ \Phi_\mu \epsilon^{\mu \nu \lambda \rho} \partial_\nu G_{\lambda \rho}
\end{equation}
Here $G_{\mu \nu}$ is a two form gauge field, with transformation 
$\delta G_{\mu \nu} ={\partial_\mu \epsilon_\nu -\partial_\nu 
\epsilon_\mu}$. Repeating the procedure of integrating over $G_{\mu \nu}$ 
and  $A_\mu$ which are again Gaussian, the following Lagrangian is 
obtained.
\begin{equation}
\frac{1}{2 (3!)} H_{\mu \nu \lambda}^2  + ( m B_{\mu \nu} - \Phi_{\mu \nu})^2
\end{equation}
where $\Phi_{\mu \nu} =  (\partial_\mu \Phi_\nu - \partial_\nu \Phi_\mu)$.
This again has invariance under $\delta B_{\mu \nu} = (\partial_\mu 
\epsilon_\nu - \partial_\nu \epsilon_\mu)$ and $\delta \Phi_\mu 
=  m \epsilon_\mu + \partial_\mu \chi$.

This Stuckelberg type action for $2$-form field (14), was
constructed and studied earlier by Aurilia and Takahashi. This
also describes a massive spin-one field, as can be seen in the
gauge where $\Phi_\mu$ is zero when it becomes Takahashi-Palmer
equation for spin-$1$ [21].

The interesting aspect about the actions (1) and (6) is that
both are obtainable from $5d$ theories. The $5d$ theory which
gives rise to the action (1) is the topologically massive theory
formed out of Kalb-Ramond field in $5d$, wherein, only zero-mode
is kept upon dimensional reduction in $4d$ [18].  Whereas, $5d$
Maxwell theory, upon dimensional reduction, keeping non-zero
modes give the Stuckelberg action [19]. Thus different $5d$
gauge invariant theories give $4d$ theories which are dually
related.  Interestingly {\it zero-mode} of topologically massive
$5d$ Chern-Simons theory is seen to be dual of {\it non-zero}
mode of Maxwell theory in $5d$.  The corresponding dual relation
between the two $5d$ theories, if any, is not clear.

Section 2:

Next we consider the $2+1$ dimensional Maxwell Chern-Simons theory, whose 
Lagrangian is
\begin{equation}
{\sl L} = -\frac{1}{4 g^2} F_{\mu \nu}^2  +  \theta A_\mu 
\epsilon^{\mu \nu \lambda} \partial_\nu A_\lambda 
\end{equation}
Apart from the usual gauge invariance, $\delta A_\mu = \partial_\mu 
\Lambda$, there is also a global symmetry, 
\begin{equation}
 A_\mu \rightarrow  A_\mu + \epsilon_\mu
\end{equation}
upto surface terms. Instead of gauging this global symmetry directly, it
is needed to first linearize the Chern-Simons term as 
\begin{equation}
A_\mu \epsilon^{\mu \nu \lambda} \partial_\nu A_ \lambda = 
P_\mu \epsilon^{\mu \nu \lambda} \partial_\nu A_\lambda 
-\frac{1}{4} P_\mu \epsilon^{\mu \nu \lambda} \partial_\nu P_\lambda
\end{equation}
where $P_\mu$ is an auxiliary vector field. This action(15) with
(17) used for Chern-Simon term has still the global symmetry in
(16). Also the auxiliary vector field $P_\mu$ has a local gauge
invariance $\delta P_\mu =\partial_\mu \chi$.

Now gauging the global symmetry, using a $2$-form gauge field, the action is 
given by
\begin{equation}
{\sl L} = -\frac{1}{4 g^2} (F_{\mu \nu} - B_{\mu \nu})^2  +  \frac{1}{2}
\theta P_\mu \epsilon^{\mu \nu \lambda}(F_{\nu \lambda}-B_ {\nu \lambda})
- \frac{1}{4}  \theta P_\mu \epsilon^{\mu \nu \lambda} \partial_\nu P_\lambda
\end{equation}
Adding the constraint, which makes the field strength of $B_{\mu \nu}$ flat,
the final action is
\begin{equation}
{\sl L} = -\frac{1}{4 g^2}(F_ {\mu \nu} - B_{\mu \nu})^2 
+ {{\theta} \over 2} P_\mu 
\epsilon^{\mu \nu \lambda}(F_{\nu \lambda} - B_{\nu \lambda})
- \frac{1}{4}  \theta P_\mu \epsilon^{\mu \nu \lambda} \partial_\nu 
P_\lambda + \Phi \epsilon^{\mu \nu \lambda} \partial_ \mu B_{\nu \lambda}
\end{equation}
Where $\Phi$ is the Lagrange multiplier field. 
Note that to maintain gauge invariance of $P_\mu$, the multiplier field also 
undergoes corresponding transformation $\delta \Phi = - \frac{1}{2} \theta 
\chi$. Integrating the$B_{\mu \nu}$ and $A_{\mu}$ field we get the 
Lagrangian
\begin{equation}
{\sl L} = 2 g^2 (\partial_\mu \Phi + \frac{1}{2} \theta P_\mu)^2 
-\frac{1}{4} \theta P_\mu \epsilon^{\mu \nu \lambda} \partial_\nu 
P_\lambda.
\end{equation}
By redefining 
$P^{\prime}_\mu =(\partial_\mu \Phi + \frac{1}{2}  \theta P_\mu)$ we get 
\begin{equation}
{\sl L} = 2 g^2 P^{\prime 2}_{\mu} - {1 \over {\theta}} P^{\prime}_\mu 
\epsilon^{\mu \nu \lambda} \partial_\nu  P^{\prime}_\lambda
\end{equation}
This is the self-dual Lagrangian to Maxwell-Chern-Simons theory due to 
Deser-Jackiw, who obtained it using Legendre transformation. Note that 
$g^2$ and $\theta$ have appeared as a reciprocal as in that of (15). This 
Lagrangian is, thus shown to be obtained by the usual duality procedure.

Conclusion:

In this paper we have shown that two massive gauge invariant
theories, Stuckelberg formulation and topologically massive
`$B^{\wedge}F$' theory are dually equivalent.
Maxwell-Chern-Simon theory in $2+1$ dimensions, which has a
global shift symmetry of the gauge field, is used to obtain
Deser-Jackiw model as its dual theory. It has been argued by
Aurilia and Takahashi [7], that the `$B^{\wedge}F$' theory
(called as gauge mixing mechanism by these authors) is the four
dimensional analog of Bosonised Schwinger model known in $1+1$
dimensions. Hence it should be interesting to see if the
Schwinger mechanism has a duality relation to Higgs mechanism.
Both these theories defined by the lagrangian (1) and (6) have
as massless limit, uncoupled massless spin-$1$ field and
massless spin-$0$ field. The latter is described in (1) by
Kalb-Ramond gauge field and in (6) by scalar field. Thus the
distinction between the massive gauge invariant spin-$1$
theories (1) and (6), seems to be that in the former, the
massless spin-$0$ field `eaten' by the massless vector field is
described by Kalb-Ramond gauge field and in the latter by scalar
field. It is well-known that massless Kalb-Ramond description of
spin-$0$ particle is dual to that of scalar field description.
Hence this difference in description appears as duality
equivalence.

Since, both these theories, are obtained as dimensional
reduction of different $5d$ theories, {\it i.e.}, one
topologically massive Kalb-Ramond theory [18] and other Maxwell
theory [19], with the former having zero-modes only and the
other non-zero modes only, the equivalence shown may have some
implication for the $5d$ theory. It should be interesting to
investigate the relation, if any, between the two $5d$ theories.
Since the $2+1$ dimensional version of (14), with two vector
field, instead of a vector and anti-symmetric tensor in $3+1$
dimension, has been shown recently to be a realization of
Josephson junction arrays [5], the equivalence shown may have an
implication there also. Also there has been generalization to
higher spin theories of the Stuckelberg formulation, it should
be interesting to obtain generalization of such topologically
massive theories to higher spin fields also. It should also be
interesting to extend this method of obtaining dual theory, also
to non-abelian `$B^{\wedge}F$' theory [20]. Work along these
lines is in progress.

Acknowledgement: We thank Prof. V. Srinivasan for encouragement.

\noindent{\bf References}
\begin{enumerate}
\item J. Schwinger, {\it Phys. Rev.} {\bf 125}, 397 (1962); ibid {\bf 128}, 
  (1962) 2425.
\item E.C.G. Stuckelberg, {\it Helv. Phys. Acta.} {\bf 11}, (1938) 225.
\item M. Kalb and P. Ramond, {\it Phys. Rev.} {\bf D9}, (1974) 2273.\\
 S. Deser and E. Witten, {\it Nucl. Phys.} {\bf B178}, (1981) 491.
\item E. Cremmer and J. Scherk, {\it Nucl. Phys.} {\bf B72}, (1971) 117.
\item P. Sodano {\it et al.,} {\it Nucl. Phys.} {\bf B448}, (1995) 505. 
{\bf CERN} {preprint, hep-th/9511168}.        
\item Theodore J. Allen, M. J. Bowick and A.Lahiri, {\it Mod. Phys. Lett.}
 {\bf A6}, (1991) 559.\\
 M. Leblanc, R. MacKenzie, P.K. Panigraghi, R.Ray, {\it Int. J. Mod. Phys.} 
{\bf A9}, (1994) 4717.\\
A.P. Balachandran and P. Teotonio-Sobrinho, {\it Int. J. Mod. Phys.} 
{\bf A9}, (1994) 1569.
\item A. Aurilia and Y. Takahashi, {\it Prog. Theo. Phys.} {\bf 66}, (1981)
 693 ; {\it Phys. Rev.} {\bf D23}, (1981) 752.
\item S.D. Rindani and M. Sivakumar, {\it Phys. Rev.} {\bf D32}, 
(1984) 3235.
\item S.D. Rindani and M. Sivakumar, {\it Z. Phys.} {\bf C49}, 
(1991) 601.\\  M. Sivakumar, {\it Phys. Rev.} {\bf D37}, (1989) 1690.
\item W. Siegel and Zweibech, {\it Nucl. Phys.} {\bf B263}, (1985) 105.
\item A. Gireon, M. Porrati and E. Rabinovici, {\it Phys. Rep.} {\bf 224},
 (1987)  77. \\
 A. Sen, {\it Int. J. Mod. Phys.} {\bf A8}, (1993) 5079.
\item T.H. Buscher, {\it Phys. Lett.} {\bf A8}, (1987)  51 ;ibid 
{\bf B201}, (1988) 466.
\item M. Rocek and E. Verlinde, {\it Nucl. Phys.} {\bf B373}, (1992) 630.
\item C.P. Burgess and F. Quevedo, {\it Nucl. Phys.} {\bf B121}, (1994) 173.
\item E. Witten, {\it On S-Duality in Abelian Gauge Theories} 
(hep-th/9505186).
\item E. Alvarez, L. Alavrez-Gaume and Y. Lozano, {\it Phy. Lett.} 
{\bf B336}, (1994) 183; ibid {\bf B355}, (1995) 165.
\item S. Deser and R. Jackiw, {\it Phys. Lett.} {\bf B139}, (1984) 371.\\
  S. K. Paul and A. Khare, {\it Phys. Lett.} {\bf B171}, (1986) 244.
\item T.R. Govindarajan, {\it J. Phys} {\bf G8}, (1982) 117.
\item  T.R. Govindarajan, S.D. Rindani and M. Sivakumar,{\it Phys. Rev.}
 {\bf D32}, (1984) 454.
\item D.Z. Freedman and P.K. Townzend, {\it Nucl. Phys.} {\bf B177},
 (1981) 282.\\
 A.P. Balachandran {\it et al.,} {\it Phys. Rev.} {\bf D26}, (1982) 1433.
\item Y. Tkahashi and Palmer,{\it Phys. Rev.} {\bf D1}, (1970) 2974.
\end{enumerate}
\end{document}